\documentclass[letterpaper, 11pt]{llncs}
\usepackage{times}
\usepackage[T1]{fontenc}
\usepackage{amsmath}
\usepackage{amssymb}
\usepackage{fancybox}
\usepackage{graphicx}
\usepackage{epsfig}
\newlength{\boxwidth}
\begin{document}
\title{Near approximation of maximum weight matching
through efficient weight reduction}
\author{Andrzej Lingas
\inst{1}
\and
Cui Di
\inst{1}
\institute{
Department of Computer Science, Lund University, 22100 Lund, Sweden.
\texttt{Andrzej.Lingas@cs.lth.se} \texttt{dcdcsunny@gmail.com} 
}
}
\date{}
\maketitle
\begin{abstract}
Let $G$ be an edge-weighted hypergraph on $n$ vertices, $m$ edges
of size $\le s,$
where the edges have real weights in an interval $[1,\ W].$
We show that if we can 
approximate a maximum
weight matching in $G$ within factor $\alpha$
in time $T(n,m,W)$ then
we can find a matching of weight at least
$(\alpha-\epsilon)$ times the maximum weight of a matching
in $G$ in time $(\epsilon^{-1})^{O(1)}\times$ \\
$\max_{1\le q \le O(\epsilon 
\frac {\log {\frac n {\epsilon}}} {\log \epsilon^{-1}})}
\max_{m_1+...m_q=m}\sum_1^qT(\min\{n,sm_j\},m_{j},(\epsilon^{-1})^{O(\epsilon^{-1})}).$
In particular, if we combine our result with the recent
$(1-\epsilon)$-approximation algorithm 
for maximum weight matching in graphs 
due to Duan and Pettie whose time complexity
has a poly-logarithmic dependence on $W$
then we obtain a
$(1-\epsilon)$-approximation algorithm for maximum weight
matching in graphs running in time 
$(\epsilon^{-1})^{O(1)}(m+n).$
\end{abstract}
\section{Introduction}

A {\em hypergraph} $G$ consists of a set $V$ of vertices 
and a set of subsets of $V$ called edges of $G.$ 
In particular, if all the edges are of cardinality
two then $G$ is a graph. A {\em matching}
of $G$ is a set of pairwise non-incident edges
of $G.$ If real weights are assigned to the edges
of $G$ then a {\em maximum weight matching} of $G$
is a matching of $G$ whose total weight achieves
the maximum. 

The problem of  finding a maximum weight matching
in a hypergraph is a fundamental generalization of 
that of finding maximum cardinality matching
in a graph. The latter is one of the basic difficult
combinatorial problems that still admit polynomial-time
solutions. For hypergraphs the decision version
of the maximum weight matching problem is NP-hard
even if the edges are of size $O(1)$ 
since it is a generalization of the problem of
maximum weight independent set for
bounded degree graphs \cite{Ho97}.
On the other hand, polynomial-time algorithms
yielding $(d-1+1/d)$-approximation of maximum weight
matching in hypergraphs with edges of size $d$
are known \cite{CL}.

The fastest known algorithms for maximum weight
matching in graphs have
substantially super-quadratic
time complexity in terms of the number $n$ 
of vertices of the input graph $G$ \cite{Ga02,Ga03,S}.
For these reasons, there is a lot of interest
in designing faster approximation algorithms
for maximum weight matching \cite{DS,DS03,DS0303,HH,PS,Pr}.

Recently, even fast approximation schemes
for maximum weight matching in graphs
have been presented
\footnote{In a preliminary version of this paper
presented at SOFSEM Student Forum held in January 2010 (no proceedings),
an $O(n^{\omega}\log n)$-time approximation
scheme for maximum weight matching in bipartite graphs
has been presented.}. The fastest known in the
literature is due to Duan and Pettie \cite{DP10}.
It yields a $(1-\epsilon)$-approximation
in time $O(m\epsilon^{-2}\log^3n )$ for a connected graph
on $n$ vertices and $m$ edges
with real edge weights. The approximation scheme
from \cite{DP10} is a composition of a 
$(1-\epsilon)$-approximate reduction of 
the problem in general edge weighted
graphs to that in graphs with small edge weights 
and an efficient $(1-\epsilon)$-approximate algorithm
for graphs with small edge weights.


\subsection{Our contributions}

Let $G$ be an edge-weighted hypergraph on $n$ vertices, $m$ edges
of size $\le s,$
where the edges have size real weights in an interval $[1,\ W].$
We show that if we can 
approximate a maximum
weight matching in $G$ within factor $\alpha$
in time $T(n,m,W)$ then
we can find a matching of weight at least
$\alpha - \epsilon $ times the maximum weight of a matching
in $G$ in time $(\epsilon^{-1})^{O(1)}\times $\\ 
$\max_{1\le q \le O(\epsilon 
\frac {\log {\frac n {\epsilon}}} {\log \epsilon^{-1}})}
\max_{m_1+...m_q=m}\sum_1^qT(\min\{ n,sm_j\},m_{j},(\epsilon^{-1})^{O(\epsilon^{-1})}).$

This reduction of maximum weight matching 
in hypergraphs with arbitralily large edge weights to that
in hypergraphs with small edge weights is incomparable
to the aforementioned similar
reduction for graphs from \cite{DP10}.
In particular, if we combine our reduction with the 
aforementioned
$(1-\epsilon)$-approximation algorithm 
for maximum weight matching in graphs 
from \cite{DP10} whose time complexity
has a poly-logarithmic dependence on $W$ then we obtain 
a $(1-\epsilon)$-approximation algorithm for maximum weight
matching in graphs running in time 
$(\epsilon^{-1})^{O(1)}(m+n).$
In comparison with the approximation scheme from \cite{DP10},
our approximation scheme is more truly linear in $m+n$,
as free from the poly-logarithmic in $n$ factor
at the cost of larger polynomial dependence on $\epsilon^{-1}.$ 
  
As another corollary from our approximate edge-weight reduction for
hypergraphs, we obtain also some results on approximating maximum
weight independent set in graphs of bounded degree.

\subsection{Other related results}

As the problem of finding maximum weight matching in graphs is a
classical problem in combinatorial optimization there is an extensive
literature on it. It includes such milestones as an early algorithm of
Kuhn \cite{K} just in the bipartite case
and an algorithm of Edmond and
Karp \cite{EK} running in time $O(nm^2)$, where $n$ is the number of
vertices and $m$ is the number of edges in the input graph.
Hungarian algorithm \cite{K} can be implemented in time $O(mn+n^2\log
n)$ with the help of Fibonacci heaps \cite{FT}
and this upper bound can be
extended to include general graphs \cite{Ga01}.

Assuming integer edge weights in $[-W,W]$ and
RAM model with \\
$\log (\max \{N,n\})$-bit words,
Gabow and Tarjan established
$O(\sqrt{n}m\log(nW))$ and $O(\sqrt{n\log n}m\log(nW))$ time-bounds
for maximum weight matching
respectively in bipartite and general graphs
\cite{Ga02,Ga03}. 

More recently, Sankowski designed an
$O(n^{\omega}W)$-time algorithm for the\\
weighted matching problem in bipartite graphs with
integer weights,
where $\omega$ stands for the exponent
of fast matrix multiplication known to not exceed $2.376$
\cite{S}. His result asymptotically improved
an earlier upper-time bound for maximum weight matching
in bipartite graphs with integer weights of the form $O(\sqrt nmW)$
due to Kao \cite{Kao}. 

There is also an extensive literature on fast approximation
algorithms for maximum weight matching in graphs 
\cite{DS,DS03,DS0303,HH,PS,Pr}. Typically they yield
an approximation within a constant factor between
$\frac 12$ and almost $\frac 45,$ running in time
of order $m\log^{O(1)}n.$ Already the straightforward
greedy approach yields $\frac 12$-approximation
in time $O(m\log n).$

The maximum weight matching problem in hypergraphs
is known also as a set packing problem in combinatorial
optimization \cite{Ho97}. By duality it is equivalent to maximum
weight independent set and hence extremely hard
to approximate in polynomial time \cite{H99}.
The most studied case of maximum weight matching
in hypergraphs is that for $d$-uniform hypergraphs
where each edge is of size $d.$  Then a polynomial-time
$(d-1+1/d)$-approximation is possible \cite{CL}. 
By duality, one obtains also a
polynomial-time $(d-1+1/d)$-approximation
of maximum weight independent set in graphs of
degree $d$ (cf. \cite{Ho97}).

\section{Simple edge weight transformations}

In this section, we describe two simple transformations
of the edge weights in the input hypergraph $G$ such that
an $\alpha$-approximation of  maximum weight matching
in the resulting hypergraph yields an
$(\alpha-\epsilon)$-approximation
of maximum weight matching of $G.$ We assume w.l.o.g. throughout
the paper that $G$ has $n$ vertices, $m$ edges,
and real edge weights not less than $1.$
The largest edge weight in $G$ is denoted by $W.$

\begin{lemma}\label{lem: max}
Suppose that there is an $\alpha$-approximation
algorithm
for maximum weight matching in $G$
running in time $T(n,m,W)$. Then,
there is an $O(n+m)$-time transformation
of $G$ into an isomorphic
hypergraph $G^*$ with edge weights in the
interval $[1, \frac {n} {\epsilon}]$
such that the aforementioned algorithm
run on $G^*$
yields an $(\alpha-\epsilon)$-approximation
of maximum weight matching in $G$
in time $T(n,m,\frac {n} {\epsilon} )$.
\end{lemma}

\begin{proof}
We may assume w.l.o.g that $W> \frac {n} {\epsilon}.$
Note that the total weight of maximum weight
matching in $G$ is at least $W.$
Hence, if we transform $G$ to a hypergraph
$G'$ by raising the weight of all edges
in $G$ of weight smaller than $\frac {W\epsilon} {n}$
to $\frac {W\epsilon} {n} $ then
the following holds:

\begin{enumerate}
\item the maximum weight of a matching in $G'$ is not less
than that in $G;$

\item any matching in $G'$ induces a matching
in $G$ whose weight is smaller by at most ${\epsilon}W.$
\end{enumerate}

To find an $\alpha$-approximation
of maximum weight matching in $G',$
we can simply rescale the edge weights in $G'$
by multiplying them by $\frac {n} {W\epsilon}.$
Let $G^*$ denote the resulting graph.
Now it is sufficient to run the asumed
algorithm on $G^*$ to obtain an
$(\alpha-\epsilon)$-approximation of maximum
weight matching in $G.$
Note that the application of the algorithm
will take time $T(n,m, \frac {n} {\epsilon}).$
\qed
\end{proof}

\begin{lemma}\label{lem: round}
Suppose that there is an $(\alpha-\epsilon)$-approximation
algorithm
for maximum weight matching in $G$
running in time $T'(n',m', W' , \epsilon )$.
By rounding down each edge weight
to the nearest power
 of $1+\epsilon$ and then running
the $(\alpha-\epsilon)$-approximation
algorithm
on the resulting graph, we obtain
an $(\alpha-O(\epsilon))$-approximation
of maximum weight matching in $G$ in time
$T'(n,m,W, \epsilon )+O(n+m).$
\end{lemma}

\begin{proof}
Let $e$ be any edge in $G.$ Denote its weight
in $G$ by $w(e)$ and its weight in the resulting
graph by $w'(e).$ We have $w'(e)(1+\epsilon)\ge w(e).$
Consequently, we obtain $w'(e)\ge w(e)-\epsilon w'(e)
\ge (1-\epsilon)w(e).$ It follows that a
maximum weight matching 
in the resulting graph has weight
at least $1-\epsilon$ times 
the weight of a maximum weight
matching in $G.$ Thus, if we run the asumed 
$(\alpha -\epsilon)$-approximation algorithm
on the resulting graph then the produced matching
with edge weights restored back to their original
values will yield an $(\alpha-2\epsilon)$-approximation.
\qed
\end{proof}


\section{A transformation into an $(\alpha-\epsilon)$-approximation
algorithm}

A {\em subhypergraph} of a hypergraph $H$ is any hypergraph
that can be obtained from $H$ by deleting some vertices
and some edges. A class $C$ of hypergraphs such that
any subhypergraph of a hypergraph in $C$
also belongs to $C$ is called {\em hereditary}.

In this section, we present a transformation of
a hypothetic $\alpha$-approximation algorithm for maximum weight matching
in a hereditary family of hypergraphs 
with edges of size $O(1)$ into a $(\alpha-\epsilon)$-approximation
algorithm. The running time of the $(\alpha-\epsilon)$-approximation
algorithm
is close to that of the $\alpha$-approximation algorithm in case
the largest edge weight is $\epsilon^{-O(\epsilon^{-1})}.$

\begin{theorem} \label{theo: main}
Suppose that there is
an algorithm for a maximum
weight matching in any hypergraph 
having edges of size $\le s$ and
belonging to the same hereditary
class as $G$
running in time
$T(n',m',W')=\Omega (n'+m'),$
where $n'$, $m'$ are respectively the
number of vertices and edges, and $[1,W']$
is the interval to which all edge weights belong.
There is an $(\alpha-\epsilon)$-approximation
algorithm for
a maximum weight matching in $G$ running in time $(\epsilon^{-1})^{O(1)}\times $\\
$\max_{1\le q \le O(\epsilon 
\frac {\log {\frac n {\epsilon}}} {\log \epsilon^{-1}})}
\max_{m_1+...m_q=m}\sum_1^qT(\min\{n,sm_j\},
m_{j},(\epsilon^{-1})^{O(\epsilon^{-1})}).$
\end{theorem}

\begin{proof}
We may assume w.l.o.g that $W=O(n/\epsilon)$
and any edge weight is a nonnegative integer
power of $1+\epsilon$
by Lemmata \ref{lem: max}, \ref{lem: round}.
Order the values of the edge weights in $G$
in the increasing
order. Set $k=O(\epsilon^{-1})$
and $l=\lceil \log_{1+\epsilon}\frac 2 {\epsilon} \rceil .$
By the form of the edge weights and the setting of $l,$
the following holds.
\par
\vskip 3pt
\noindent
{\bf Remark 1}: For any two different edge weights
$w_1$ and $w_2,$ if the number of $w_1$ is greater than
that of $w_2$
 by at least $l$
in the aforementioned
ordering then $\frac {\epsilon}2 w_1\ge w_2.$
\par
\vskip 3pt
In order to specify our $(\alpha-\epsilon)$-approximation
algorithm, we partition
the ordered edge weights into consecutive closed basic intervals,
each but perhaps for the last, containing exactly $l$
consecutive edge weights, see Fig. \ref{fig: basic}.

\begin{figure}[h]
\begin{center}
  \includegraphics[width=12cm]{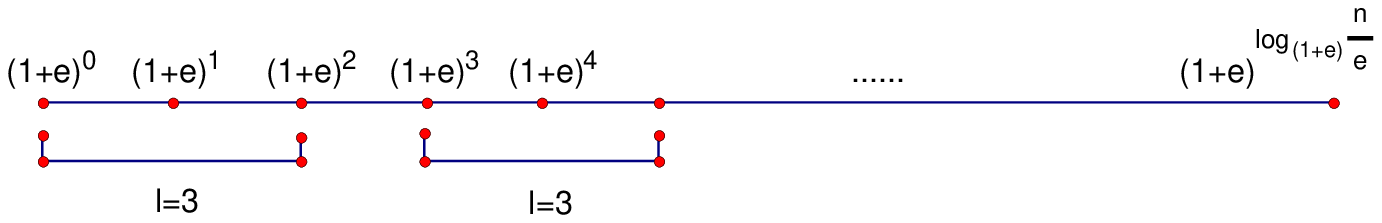}\\
  \caption{Partitioning of edge weights ($l=3$)}
\label{fig: basic}
\end{center}
\end{figure}

 Next, we group $k$-tuples of consecutive
basic intervals into large intervals composed of $k-1$ consecutive
basic intervals followed by a single basic interval
called a gap. This partition
corresponds to the situation
when the so called shift parameter $x$ is set to $0.$
For $x\in \{1,..,k-1\},$ the partition into alternating
large intervals and gaps is shifted by $x$ basic intervals
from the right, so the first large interval from
the right is composed solely of $k-1-x$ basic intervals,
see Fig. \ref{fig: shift}. The maximal subgraph of $G$
containing solely edges in the large intervals
in the partition is denoted by $G_x.$

\begin{figure}[h]
\begin{center}
  \includegraphics[width=12cm]{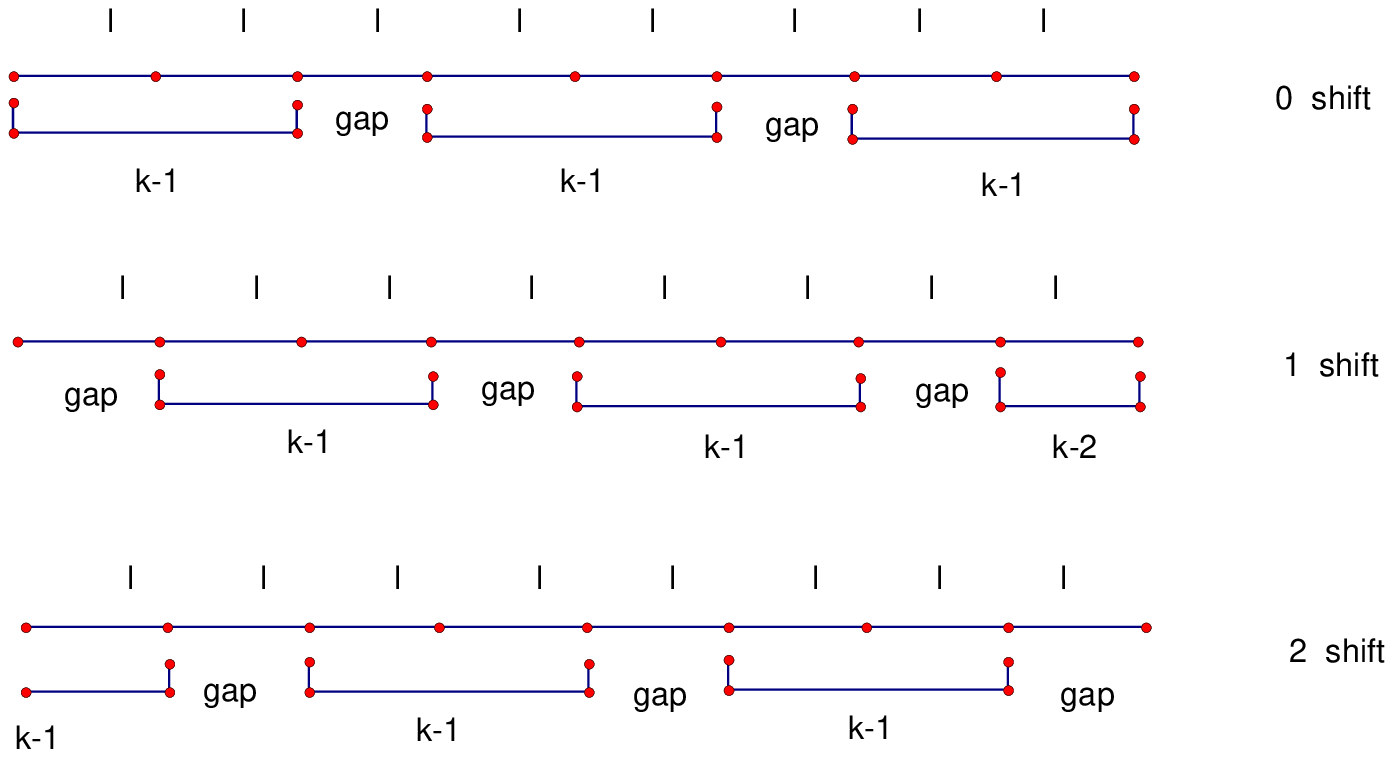}\\
  \caption{An example of shift: l=3, k=3}
\label{fig: shift}
\end{center}
\end{figure}

For our $(\alpha-\epsilon)$-approximation 
algorithm for a maximum weight matching
in $G$ see Fig. \ref{fig: alg}.
We shall assume the definitions
of the subgraphs $G'_x,G_{x,j},M_x$
given the algorithm.

\begin{figure}
\noindent
\label{fig: alg}
{\bf Algorithm 1}

\begin{enumerate}
\item
\label{p-1}
{\bf for} $x\leftarrow 1$ {\bf to} $k-1$ {\bf do}
\item
\label{p-2}
\hskip 0.5cm $M_{x}\leftarrow \emptyset;$
\item
\label{p-3}
\hskip 0.5cm $G'_{x} \leftarrow G_{x}$;
\item
\label{p-4}
\hskip 0.5cm {\bf for} $j\leftarrow 1$ {\bf to} $O(\log_{1+\epsilon}\frac
n {\epsilon})$ {\bf do}
\item
\hskip 0.5cm {\bf begin}
\item
\hskip 1cm  Set $G_{x,j}$ to
the sub-hypergraph of $G'_{x}$ induced by the edges whose weights
\item
\hskip 1cm
fall in the $j$th interval from the right;
\item
\hskip 1cm Run the $\alpha$-approximation
algorithm for maximum weight matching
$M_{x,j}$ of $G_{x,j}$;
\item
\hskip 1cm $M_{x}\leftarrow M_{x} \bigcup M_{x,j}$;
\item
\hskip 1cm Remove all edges incident to $M_{x,j}$ from $G'_{x}$;
\item
\hskip 0.5cm {\bf end}
\label{p-5}
\item
Return the heaviest among the matchings $M_x$
\end{enumerate}
\caption{The $(\alpha -\epsilon)$-approximation algorithm.}
\end{figure}

Since the union of the gaps over all shifts covers
all weights there must a shift where the gaps
cover at most $\frac 1k$ of the weight of optimal
matching of $G.$ Hence,
there must be a shift $x$ such that
the weight of optimal matching in $G_x$ is at least
$(1-1/k)$ of the weight of optimal matching of $G.$
Thus, it is sufficient to show that $M_x$ closely
approximates an $\alpha$-approximate weight matching of $G_x.$

Consider a maximum weight matching $OM_x$ of $G_x$
and the $\alpha$-approximation $M_{x.j}$
of a maximum weight matching of $G_{x,j},$
respectively. Note that $M_{x,j}$ has total weight
not smaller than $\alpha$ times
the total weight of $OM_x$ restricted to
the edges in $G_{x,j}.$ On the other hand,
each edge $e$ in $M_{x,j}$ can eliminate at most $O(1)$
edges of $OM_x$
from all $G_{x,i}$ for $i>j.$ The total weight of the at most
$O(1)$ edges is only at most the $\epsilon$ fraction
of the weight of $e$ by Remark 1. Let $EOM_x$ denote
the set of all edges in $OM_x$ eliminated by $M_x=\bigcup_jM_{x,j}$.
The following two inequalities follow:

$$weight(M_x) + weight(EOM_x)\ge \alpha \times weight(OM_x)$$
$$\epsilon \times weight(M_x)\ge weight(EOM_x)$$

Consequently, we obtain:

$$weight(M_x)\ge \alpha\times
weight(OM_x)-\epsilon \times weight(M_x)\ge (\alpha -\epsilon)\times
weight (OM_x)$$


Thus, $M_x$ approximates within $(\alpha-\epsilon )$
a maximum weight matching of $G_x,$ and
consequently the heaviest of the matchings $M_x$
approximates within $(1-\epsilon)(1-1/k)$
a maximum weight matching of $G.$
By setting $k=\Omega (\frac 1 \epsilon)$, we obtain
an $(1-O(\epsilon))$-approximation of the optimum.

It remains to estimate the time complexity
of our method. Note that the weight of heaviest
edge in $G_{x,j}$ is at most

$$(1+\epsilon)^{lk} = O(\epsilon^{-1})^{O(\epsilon^{-1})}
=(\epsilon^{-1})^{O(\epsilon^{-1})}$$

\noindent
times larger than that of the lightest one.
Let $m_{x,j}$ denote the number of edges
in $G_{x,j}.$ Next, let $n_{x,j}$ denote
the number of vertices in the sub-hypergraph
of $G_{x,j}$ induced by the edges of $G_{x,j}.$
Note that $n_{x,j}\le \min\{ n,sm_{x,j}\}$
by our assumption on the size of edges in $G.$

Hence, by rescaling the weights in $G_{x,j},$
we can find $M_{x,j}$ in time\\
$T(\min\{ n,sm_{x,j}\},m_{x,j},(\epsilon^{-1})^{O(\epsilon^{-1})})$ 
for $j=1..,O(\log_{1+\epsilon}
{\frac n {\epsilon}} /lk)$ and $x=0,...,k-1.$
Note that $\log_{1+\epsilon}
{\frac n {\epsilon}} =\frac {\log {\frac n {\epsilon}}}
{\log {1+\epsilon}}=\Theta (\epsilon^{-1} \log {\frac n {\epsilon}})$
and similarly $lk=\log_{1+\epsilon}\frac 2 {\epsilon}
\Theta (\epsilon^{-1})=\Theta (\frac {\log \frac 2 {\epsilon}}
{\log {1+\epsilon}}\epsilon^{-1})=\Theta
(\epsilon^{-2}\log \epsilon^{-1}).$
It follows that for a given $x,$ the largest value of $j,$
i.e., the number of the subgraphs $G_{x,j}$ is
$O(\epsilon \frac {\log {\frac n {\epsilon}}} {\log \epsilon^{-1}}).$

Note that $\sum_j m_{x,j}\le m$ since each edge of $G$
belongs to at most one hypergraph $G_{x,j}.$
Thus, the total time taken by finding all $M_{x,j}$
for $j=1,..., O(\epsilon 
\frac {\log {\frac n {\epsilon}}} {\log \epsilon^{-1}})$
for a fixed $x$ is \\
$\max_{1\le q \le O(\epsilon 
\frac {\log {\frac n {\epsilon}}} {\log \epsilon^{-1}})}
\max_{m_1+...m_q=m}\sum_1^qT(\{n,sm_j\},m_{j},(\epsilon^{-1})^{O(\epsilon^{-1})}).$\\
Recall that $x$
ranges over
$O(\epsilon^{-1})$ possible values.

By the assumed form of the edge weights in $G,$
we can apply a standard radix sort with
$O(\epsilon^{-1} \log {\frac n {\epsilon}})$ buckets
to sort the edges of $G$ by their weights
in time $O(m+\epsilon^{-1} \log {\frac n {\epsilon}})$.
The latter is also 
$O(\epsilon^{-2}T(n,m,(\epsilon^{-1})^{O(\epsilon^{-1})}))$
by the assumptions on $T.$ 

In order to efficiently
construct the graphs $G_{x,j}$,
the sorted edge list is kept in array and there are
double links between an occurrence of an edge in the adjacency
lists representing $G$ and its occurrence
in the sorted edge list. To determine the edges
inducing $G_{x,j}$, we just
scan a consecutive fragment of the sorted list
from left to right. Given a list of edges of
$G_{x,j}$, an adjacency representation of
the sub-hypergraph can be constructed in time
$O(n+m)=O(T(n,m,(\epsilon^{-1})^{O(\epsilon^{-1})}))$ by
using the aforementioned double links.

To remove an edge from $G'_x,$
we locate it on the sorted
edge list by using the double links
with the adjacency lists and
then link its predecessor with its successor
on the sorted list. We conclude that
the updates of $G'_x$ take time 
$O(m)=$
$O(T(n,m,(\epsilon^{-1})^{O(\epsilon^{-1})}))$.
\qed
\end{proof}

\section{Applications}

There are at least two known exact algorithms for maximum
weight matching in bipartite graphs 
with integer edge weights for which
the upper time bounds on their running
time in linear fashion depend on the
maximum edge weight $W$ \cite{Kao,S}.
Recently, Duan and Pettie have provided
substantially more efficient $1-\epsilon$
approximation algorithm for maximum weight
matching in general graphs with integer
edge weights, whose running time also
depends on $W$ in linear fashion \cite{DP10}. 
Furthermore, their final approximation
scheme for this problem in fact
exhibits poly-logarithmic dependence
on $W.$
\par
\vskip 3pt
\noindent
{\bf Fact 1} (Duan and Pettie, see the proof
of Theorem 1 in \cite{DP10}). {\em An
$(1-\epsilon)$-approximation of maximum weight
matching 
in a connected graph
on $m$ edges
and positive integer weights
not exceeding $W$
can be found deterministically in time $O({\epsilon}^{-2} m\log^3 W).$}
\par
\vskip 3pt
\noindent
We can trivially generalize the upper time bound
of Fact 1 to include a non-necessarily connected
graph by extending it by an additive factor of $O(n).$

There is one technical difficulty in combining Facts 1
with Theorem \ref{theo: main}. Namely, in the theorem
we assume that there is available an 
$\alpha$-approximation algorithm
for maximum weight matching for graphs belonging
to the same hereditary class as $G$ with arbitrary
real edge weights not less than $1$ whereas the
algorithm of Facts 1 assumes
integer weights. In fact, even if the input graph
got positive integer weights the preliminary edge
weight transformations in the proof of Theorem \ref{theo: main}
would result in rational edge weights. There is a simple
remedy for this. We may assume w.l.o.g that $\epsilon$
is an inverse of a positive integer and through all
the steps of our approximation scheme round down
the edge weights to the nearest fraction with denominator
$O(\epsilon^{-1})$ and then multiply them by the common
denominator to get integer weights. This will increase
the maximum weight solely by $O(\epsilon^{-1})$ and will
preserve close approximability.

Hence, Fact 1 combined in this way
with Theorem \ref{theo: main} yield our main application result
by straightforward calculations.

\begin{theorem} \label{theo: amain}
There is an approximation scheme for
a maximum weight matching in
a graph on $n$ vertices and
$m$ edges
running in time
$(\epsilon^{-1})^{O(1)}(m+n)$.
\end{theorem}

\section{Extensions}

Note that Theorem \ref{theo: main} includes
as a special case the problem of finding
a maximum weight independent set in a
graph $G$ of maximum degree $d$ which is equivalent
to the problem of finding a maximum weight
matching in the dual hypergraph with
edges corresponding to the vertices of $G$
and {\em vice versa}.

Several combinatorial algorithms for maximum independent
set achieving
the approximation ratio of $O(d)$, where $d$ is
the maximum or average degree are known in the literature
\cite{Ho97}. In the appendix, we demonstrate that
by using the method of Theorem \ref{theo: main}
they can be simply
transformed into good approximation
algorithms for maximum weight independent set.

\section{Acknowledgments}

The authors are very grateful to Seth Pettie for his valuable
suggestions to apply the second $(1-\epsilon)$-approximation algorithm
from \cite{DP10} instead of the first one and to eliminate an $n\log
n$ term in the application in an intermediate version of our
paper. They are also very grateful to anonymous referees for valuable
comments on an early version of the paper presented solely orally at
Student Forum of SOFSEM 2010.

\vfill
\newpage
\section{Appendix: Approximation algorithms for maximum
weight independent set in bounded degree graphs}

Note that Theorem \ref{theo: main} includes
as a special case the problem of finding
a maximum weight independent set in a
graph $G$ of maximum degree $d$ which is equivalent
to the problem of finding a maximum weight
matching in the dual hypergraph with
edges corresponding to the vertices of $G$
and {\em vice versa}.

Several combinatorial algorithms for maximum independent
set achieving
the approximation ratio of $O(d)$, where $d$ is
the maximum or average degree are known in the literature
\cite{Ho97}. Here, we demonstrate that
by using the method of Theorem \ref{theo: main}
they can be simply
transformed into good approximation
algorithms for maximum weight independent set.

\begin{lemma}\label{lem:ind}
Suppose that there is an $\alpha (d)$-approximation algorithm for
maximum independent set in a graph on $n$ vertices and maximum
(or average degree, respectively) degree
$d$ running in time $S(n,d)$, where the function $S$ is non-decreasing
in both arguments. There is an
$\alpha(dW)$-approximation algorithm for maximum
weight independent set in a graph on $n$ vertices, maximum
(or average degree, respectively) degree $d,$
positive integer weights not exceeding an integer $W,$ running in time
$S(nW, dW)$.
\end{lemma}

{\bf Proof:}
Let $G$ be the input vertex weighted graph $G.$ We form the auxiliary
unweighted graph $G^*$ on the base of $G$ as follows. In $G^*$,
we replace each
vertex $v$ in $G$ with the number of its copies equal to the weight
of $v$. We connect each copy of $v$ by an edge with each copy of each
neighbor of $v.$ Next, we run the assumed algorithm for maximum
unweighted independent set on $G^*.$ Note that any maximal independent
in $G^*$ is in one-to-one correspondence with an independent set in
$G$ since whenever a  copy of $v$ is in the independent set then all
other copies of $v$ can be inserted into it without any conflicts.\qed

The drawback of Lemma \ref{lem:ind} is that
the approximation factor and/or the running time of the resulting
algorithm for the weighted case can be very large in case the maximum
weight $W$ is large. However, we can plug
Lemma \ref{lem:ind} in the method of Theorem \ref{theo: main}
to obtain much more interesting approximation algorithms in
the weighted case.

\begin{theorem}
Suppose that there is an $\alpha (d)$-approximation algorithm for
maximum independent set in a graph on $n$ vertices and maximum
degree
$d$ running in time $S(n,d)$, where the function $S$ is non-decreasing
in both arguments and $S(n,d)=\Omega (nd\log n) $. There is an
$(\alpha(d\epsilon^{-1})^{O(\epsilon^{-1})})-d\epsilon )$-approximation algorithm for maximum
weight independent set in a graph on $n$ vertices, with maximum
degree $d,$
positive integer vertex weights, running in time\\
$O(\epsilon \frac {\log (n/\epsilon)} {\log \epsilon^{-1}}S(n(\epsilon^{-1})^{O(\epsilon^{-1})}, d(\epsilon^{-1})^{O(\epsilon^{-1})}))$.
\end{theorem}

\begin{proof} sketch. 
Recall that the problem of maximum (weighted
or unweighted) independent set is equivalent to the problem
of maximum (weighted or unweighted, respectively) matching
in the dual  hypergraph. In the dual hypergraph,
the edges have size
not exceeding
the maximum vertex degree
in the input graph.
We run the method of Theorem \ref{theo: main} on the dual hypergraph
using as the black box algorithm the result
of the application of Lemma \ref{lem:ind} to the
assumed algorithm and its adaptation to the maximum
matching problem in the dual hypergraph.\qed
\end{proof}


{\small
\begin{thebibliography}{10}
\addcontentsline{toc}{chapter}{\numberline{}Bibliography}

\bibitem{B00}
P. Berman.
\newblock A  $d/2$ Approximation for
Maximum Weight Independent Set in $d$-Claw Free Graphs.
\newblock Proc. 7th SWAT, Lecture Notes in Computer
Science, Springer, Volume 1851, pp. 31-49, 2000. 

\bibitem{be}
T. H. Cormen, C. E. Leiserson, R. L. Rivest, and C. Stein.
\newblock Introduction to Algorithms.
\newblock \emph{2nd edition, McGraw-Hill Book Company, Boston, MA, 2001.}

\bibitem{CL}
Y.H. Chan and L.C. Lau.
\newblock On Linear and Semidefinite Programming Relaxations 
for Hypergraph Matching.
\newblock Proc. 

\bibitem{DS}
D. Drake and S. Hougardy.
\newblock A simple approximation algorithm for the weighted matching problem.
\newblock\emph{Info. Proc. Lett.}, 85:211-213, 2003.

\bibitem{DS03}
D. Drake and S. Hougardy.
\newblock Linear time local improvements for weighted matchings in graphs.
\newblock\emph{International Workshops on Experimental and Efficient Algorithms (WEA), LNCS 2647}, pages 107-119, 2003.

\bibitem{DS0303}
D. Drake and S. Hougardy.
\newblock Improved linear time approximation algorithms for weighted matchings.
\newblock\emph{7th International Workshops on Randomization and Approximation Techniques in Computer Science (APPROX), LNCS 2764}, pages 14-23, 2003.

\bibitem{DP10}
R. Duan and S. Pettie.
\newblock Approximating Maximum Weight Matching in Near-linear Time.
\newblock Proc. FOCS 2010.

\bibitem{EK}
J. Edmonds and R. M. Karp.
\newblock Theoretical Improvements in Algorithmic Efficiency for Network Flow Problems.
\newblock \emph{J. ACM}, 19(2):248-264, 1972.

\bibitem{FT}
M.L. Fredman and R.E. Tarjan. 
\newblock Fibonacci heaps and their uses in improved network optimization
algorithms. 
\newblock J. ACM, vol. 23, no. 2, pp. 596-615, 1987.

\bibitem{Ga01}
H. N. Gabow.
\newblock Data structures for weighted matching and nearest common ancestors with linking.
\newblock\emph{First Annual ACM-SIAM Symposium on Discrete Algorithms(SODA)}, pages 434-443, 1990.

\bibitem{Ga02}
H. N. Gabow and R. E. Tarjan.
\newblock Faster scaling algorithms for network problems.
\newblock\emph{SIAM J. Comput.}, 18(5):1013-1036, 1989.

\bibitem{Ga03}
H. N. Gabow and R. E. Tarjan.
\newblock Faster scaling algorithms for general graph-matching problems.
\newblock\emph{J. ACM}, 38(4):815-853, 1991.

\bibitem{H99}
J. Hastad.
\newblock Clique is Hard to Approximate 
within $n^{1-\epsilon}.$
\newblock Acta Math 182(1), pp. 105-142, 1999.

\bibitem{HH}
Hanke and Hougardy.
\newblock $3/4 - \epsilon$
 and $4/5 - \epsilon$
 approximate MWM algorithms running in 
$O(m log n)$ and $O(m log^2 n)$ time
\newblock University of Bonn, Research Institute for Discrete Mathematics Report No. 101010.

\bibitem{Ho97}
D. S. Hochbaum,
\newblock Approximating Covering and Packing Problems:
Set Cover, Vertex Cover, Independent Set, and Related Problems
\newblock in Approximation Algorithms for NP-hard Problems,
D.S. Hochbaum (ed.), PWS Publishing Company, Boston, 1997.

\bibitem{Kao}
M.-Y. Kao, T.-W. Lam, W.-K. Sung and H.-F.  Ting.
\newblock A Decomposition Theorem for Maximum Weight
Bipartite Matchings with Applications to Evolutionary Trees.
\newblock Proc. European Symposium on Algorithms (ESA 1999),
LNCS 1643, Springer Verlag, pp. 438-449, 1999.

\bibitem{K}
H. W. Kuhn.
\newblock The Hungarian method for the assignment problem.
\newblock\emph{Naval Research Logistics Quarterly}, 2:83-97, 1955.



\bibitem{PS}
S. Pettie and P. Sanders.
\newblock A simple linear time 2/3-$\epsilon$ approximation for maximum weight matching.
\newblock\emph{Information Processing Letters}, 91:271-276, 2004.

\bibitem{Pr}
R. Preis.
\newblock Linear time 1/2-approximation algorithm for maximum weighted matching in general graphs.
\newblock \emph{Proc. 16th Ann. Symp. on Theoretical Aspects of Computer Science (STACS), LNCS 1563}, pages 259-269, 1999.

\bibitem{S}
P. Sankowski.
\newblock Weighted bipartite matching in matrix multiplication time.
\newblock\emph{LNCS 4051}, pages 274-285, 2006.







\end{thebibliography}}
\end{document}